# A Simple Phase Retrieval Algorithm from a Single Shot Interferogram

Lifa Hu [a,b,*], Wen Shen[a,b], Wenchao Ma[a,b], Dongting Hu[a,b], and Xinyu Liu[a,b]

[a]Jiangnan University, School of Science, Department of Photoelectric
Information Science and Engineering, Wuxi, China, 214122

[b]Jiangnan University, Jiangsu Provincial Research Center of Light Industrial
Optoelectronic Engineering and Technology, Wuxi, China, 214122

**Abstract**. Traditional phase-shifting interferometry technique cannot be used to measure time-varying phase distributions. But single shot techniques could resolve the problem. Many efforts have been made on the phase retrieval methods from a single shot interferogram. In the paper, a simple and effective method is presented without complex computation. The interference fringe is transferred to a phase distribution with a look-up-table. And then it is divided into different regions according to the parity of every pixel. The pixels in the same region have the same parity, which determines the wrapped phase. Additionally, the light spot displacement of a local wavefront is obtained to solve the global sign ambiguity. The theoretical simulation results indicate that the PV of wavefront error is $0.00054\lambda$ and the rms is $0.000125\lambda$, which is much better than the results from the Fast Fourier Transformation method. We also use it in the experimentally measured interferogram. Our algorithm has the advantages of simplicity, high precision and effective for both open and closed interferometer fringes, which will be valuable for real time monitoring the optical elements' shape during their processing.

## 1 Introduction

The optical interferometry is widely used for precision measurements, surface diagnostics, astrophysics, seismology, quantum information, etc[1]. According to their principles, they can be divided into two types. One is to measure the variation of the interference fringe or the optical path difference at a specified point on the interference field to obtain the sample characteristic parameters, such as its size, displacement, material micro-deformation and refractive index. The other is to obtain surface shape, sample geometry or fluid density distribution by measuring the interference fringes generated by the measured wavefront and the reference standard wavefront. Surface shape measurement is the classic applications of the optical interferometry.

There are mainly two algorithms used to extract phase from interference fringes. The first is the phase shifting method as used in the traditional phase shifting interferometer(PSI) (see [1-4]). It requires a high-precision phase shifter based on a PZT translator to introduce a known phase shift temporally, or uses complex optical layout to generate interferograms with fixed phase steps spatially in different sub-areas of CCD. Therefore, very high phase demodulation accuracy can be obtained. However, its applications are limited in static or quasi-static situations.

In order to work in dynamic situations, real-time phase shifting interference technology has been investigated (see [5-25]). And a new algorithm different from PSI is presented to extract the phase from a single shot interferogram(SSI). And many data processing methods for SSI have been introduced, for example, wavelet analysis (see[6-9]), Hilbert transform(HT) (see[10-15]), fast Fourier transform [16-19], regularized phase tracing method (see[20-22]) and the energy minimization method (see[23-25]) and so on.

In fact, there are two sign ambiguity problems for these single shot techniques. One could be named as the local sign ambiguity(LSA) in the calculation procedure from interferogram to unwrapped phase. For an example, because the HT could not distinguish the positive and negative spatial frequencies, part of the phase will show a spurious phase jump ($\pi$ shift). The other could be named as the global sign ambiguity(GSA) in the calculation procedure from the unwrapped phase to the accurate optical surface shape. Most efforts are focused on LSA. However, to remove GSA, all of them need prior knowledge. For example, the interferometric image of a tear film on eye will show dry areas for breaking up tear film where phase is the absolutely zero (see [24]), which could be used as the prior knowledge to determine the accurate inclination direction of wavefront.

The fast Fourier transform method (FFT) is very popular in applications, which realizes the demodulation of interference fringes by adding large tilt to the interference fringe to be measured. And single shot Interferometry based on FFT is available with comercial systems in the market. However, to improve the phase extraction accuracy, reasonable parameters of the carrier frequency and window function should be selected carefully for different wavefront to be tested, which is not an easy and convenient work. In addition, it is easy to generate unreasonable phase truncation and spectral aliasing for closed fringes, which limits its application.

In the paper, we present a simple and effective method to extract phase from a single shot interferogram. The phase retrieval is mainly completed by the relationship between normalized fringe intensity and phase (named as look-up table, LUT) and the parity of every pixel. In section 2, the principle and data processing procedure of the method are presented in detail. In section 3, the accuracy of the method are evaluated theoretically and compared with those of FFT method. In section 4, experimental results are given. A conclusion is given in section 5.

## 2  Formulation of mathematic model

In the classic Michelson interferometer, the aberration of the optical surface to be measured, or its certain inclination angle relative to reference mirror, will cause the optical path difference between two reflected light beams. And the obtained interference fringes will be

bent or become dense. The intensity of interference fringe acquired by the camera can be described as

$$I = I_1 + I_2 + 2\sqrt{I_1 I_2} \cos(\Delta\phi) \tag{1}$$

Where $I_1$ and $I_2$ are the intensities of the two reflected light beams; $\Delta\phi$ is the phase difference between the measured surface and the reference one. It should be noted that considering the reflection mode of the interferometer, the phase in the cosine function is twice of the actual wavefront. In general, when the intensity of the two light beams is the same as $I_0$, equation (1) can be expressed as:

$$I = 2I_0 \left[1 + \cos(\Delta\phi)\right] \tag{2}$$

Generally, the fringes sampled with CCD in experiments are filtered and normalized to improve the sign noise ratio (SNR)[26]. According to equation (2), we define a normalized intensity $I_{norm}$ at point (x,y) as:

$$I_{norm}(x, y) = \frac{I}{4I_0} = 0.5 + 0.5 \cos\left[\Delta\phi(x, y)\right] \tag{3}$$

Where, the values of normalized intensity $I_{norm}$ obtained by equation (3) are ranged from 0 to 1. For a 8-bit CCD, we could define a linear monotonous relationship between the normalized intensity $I_{norm}$ and the gray level $I_{linear}$ as follows :

$$I_{linear}(x, y) = 255 \cdot I_{norm}(x, y) \tag{4}$$

Therefore, the phase as a function of the light intensity $I_{linear}$ according to equations (3) and (4) could be calculated as:

$$\Delta\phi_\pi(x, y) = a\cos\left[I_{norm}(x, y)\right] = a\cos\left[\frac{2}{255} I_{linear}(x, y) - 1\right] \tag{5}$$

It should be noted that the phase obtained from equation (5) is ranged from 0 to π. Therefore, for the 8-bit camera with a maximum gray level of 255, the phase as a function of the grey level corresponding to the light intensity $I_{linear}$ is shown in figure 1. The abscissa is the gray level of the fringe with 8-bit. The maximum gray value is 255 and the minimum is 0, and the phase $\Delta\phi_\pi$ in Y-axis is ranged from 0 to π. According to the monotonic relationship between phase and gray level in figure 1, a look-up table (LUT) including phase and light intensity can be built and used to compute the wrapped phase from a normalized single shot interferogram. Instead of using the arc tan function in FFT or HT methods, we use the LUT generated with the equation (5) to obtain the phase $\Delta\phi_\pi$ from normalized intensity map.

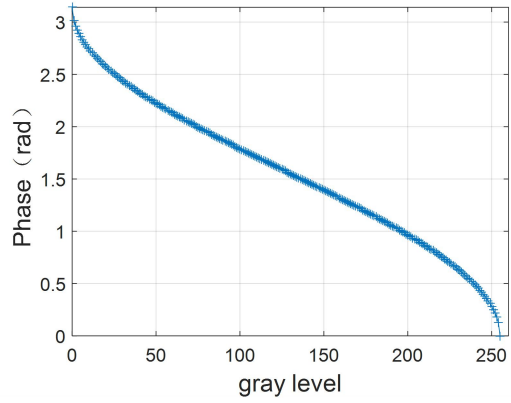

**Fig. 1** Relationship between phase and gray level.

It should be noted that nearly half parts of the phase $\Delta\phi_\pi$ in the phase map will be in a wrong direction after we directly calculate them from an interferogram according to LUT. In order to obtain the wrapped phase $\Delta\phi_{2\pi}$ from $\Delta\phi_\pi$, we could introduce an interferogram as an example. An interferogram is generated on the unit circle as shown in figure 2, its PV and RMS of wavefront are 6.23λ and 1.56λ(λ=635nm), respectively. In figure 2, the black area corresponds to the zero gray level and the white one corresponds to the gray level of 255. First, we define a matrix N that is used to record the characteristic values $n_{i,j}$ of every pixels on the interferogram, i and j are the row and column number corresponding to the coordinates of every pixels, respectively; Second, find the darkest and brightest pixels from the fringe pattern, and then set their $n_{i,j}$ to -1 and 1, respectively, and all of the rest pixel's $n_{i,j}$ are set to zero. And then the obtained maxtrix N is shown in figure 3. And the fringe pattern could be divided into different areas according to the lines in figure 3.

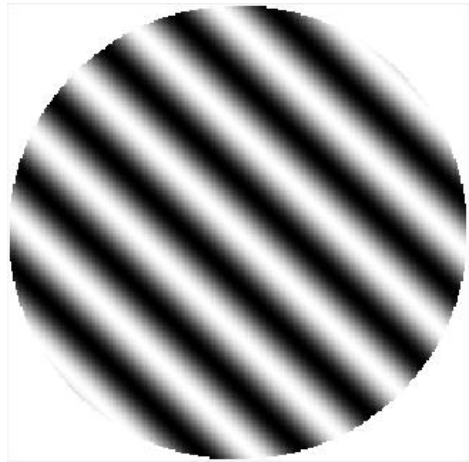

**Fig. 2** simulated interferogram due to tip and tilt.

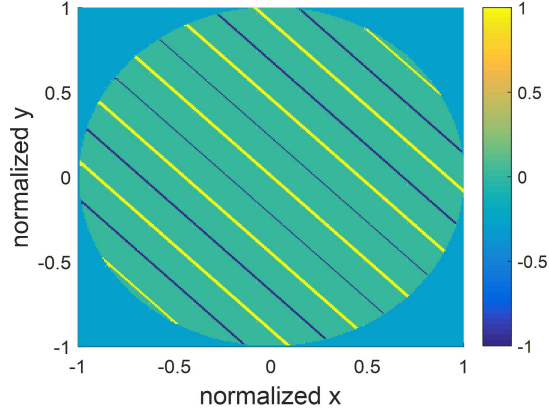

**Fig. 3** The obtained N, the brightest(yellow) and darkest(blue) part of the fringe pattern are for 1 and -1, respectively.

Then, define another matrix M and set its elements $m_{i,j}$ to every pixel in the unit circle from left to right and from bottom to top:

$$m_{i,j} = p_i + k_{i,j} \tag{6}$$

Where, $p_i$ is the value of the first pixel of every row in the unit circle. And it is equal to zero for the first pixel of the first row in the unit circle. And when it passes a line from the first row to the last row in figure 3 for the first pixel of the i-th row in the unit circle, $p_i$ increases by one. $k_{i,j}$ is zero for the first pixel of every rows in the unit circle, and then $k_{i,j}$ increases by one when a peak and valley line in figure 3 is passed from left to right. Therefore, according to equation (6), it will be obvious that $m_{i,j}$ for the pixels in the area between the adjacent peak and valley lines will have the same parity.

With the $m_{i,j}$ value of each pixel according to equation (6), phase $\Delta\phi_\pi$ could be transformed to wrapped phase $\Delta\phi_{2\pi}$ as follows:

$$\Delta\phi_{2\pi}(x, y) = (-1)^{n_{i,j}} \Delta\phi_\pi(x, y) + \pi \tag{7}$$

To obtain unwrapped phase from $\Delta\phi_{2\pi}$, a reasonable assumption is used that two adjacent points usually have a continuous phase and their phase difference is much smaller than $2\pi$. Therefore, a phase shift of $2\pi$ is added when their phase difference is larger than $\pi$. Then, it is relatively easy to obtain the unwrapped phase.

After that, we could fit the result with Zernike polynomials. The used Zernike polynomial is defined as follows[27]:

$$Z = \begin{cases} Z_{even,i} = \sqrt{n+1} R_n^m(r) \sqrt{2} \cos(m\theta), m \neq 0 \\ Z_{odd,i} = \sqrt{n+1} R_n^m(r) \sqrt{2} \sin(m\theta), m \neq 0 \\ Z_{even,i} = \sqrt{n+1} R_n^0(r), m = 0 \end{cases} \tag{8}$$

$$R_n^m(r) = \sum_{s=0}^{(n-m)/2} \frac{(-1)^s (n-s)!}{s![(n+m)/2 - s]![(n-m)/2 - s]!} r^{n-2s}, m \neq 0$$

Where both the radial order $n$ and the angular order $m$ are integers, and $m \leqq n$ and ($n$-|$m$|) is even. Arbitrary wavefront φ can be expanded with Zernike polynomials:

$$\varphi = \sum_i C_i \cdot Z_i \qquad (9)$$

Where $C_i$ is the coefficient of the i-th Zernike mode $Z_i$. Considering that it is the reflective mirror, the accurate surface should be half of the measured wavefront.

Now the only problem is GSA, and we could use a part of the reconstructed wavefront to estimate its PSF, and calculate its weight of centers along x and y axis. After that, compare the results with the measured one so as to determine the wavefront's inclination direction, which is used as the prior knowledge to solve GSA problem. In the simulation, we could select a small circular area located on the up left corner. Different global tilt and tip in the selected circular area will lead to different displacements of estimated PSF's centers along x and y, which could be used as the prior knowledge to solve GSA problem. Experimentally, we could introduce a hole and a lens in classic Michelson interferometer and remove its reference mirror as shown in figure 4. If there are displacement direction differences between the gravity centers of the measured one and the estimated one, the local tilt and tip of the selected part in estimated wavefront is in a wrong direction. And then, the right direction of wavefront could be determined.

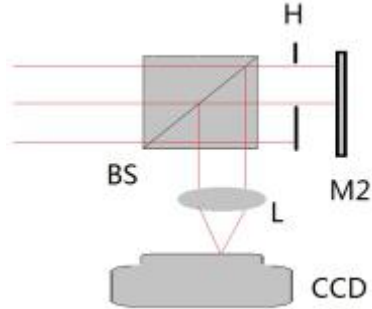

**Fig.4** Calibration optical layout: M2: mirror to be measured; BS: beam splitter; H: hole; L: lens.

### 3  Simulation and discussion

In the section 2, we use the tilt and tip terms as an example to introduce the principle data processing procedure in detail. In this section, we verify the presented method with a simulation. To evaluate the restoring precision from a single shot fringe, we calculate the residual wavefront $\Delta\varphi_{err}$ as:

$$\Delta\varphi_{err} = \Delta\varphi_{origin} - \Delta\varphi_{meas} \qquad (11)$$

Where $\Delta\varphi_{origin}$ and $\Delta\varphi_{meas}$ are the original wavefront and the measured one, respectively.

Instead of pure tilt and tip terms, we use another aberration dominated by tilt-tip and coma as shown in figure 5. Figure 5(a) shows the single shot fringe corresponding to the aberration, and figure 5(b) shows the calculated value distribution of $m_{i,j}$. For every pixel on the fringe pattern as shown in figure 5(a), its $m_{i,j}$ is calculated according to equation (6) from left to right and from bottom to top as described in section 2. It is obvious that the parity of $m_{i,j}$ is the same in the same region, that is ,they are odd or even in the same area. We have conducted a number of fringe pattern simulations. This parity property is suitable for their fringe patterns, such as closed and /or open fringe patterns.

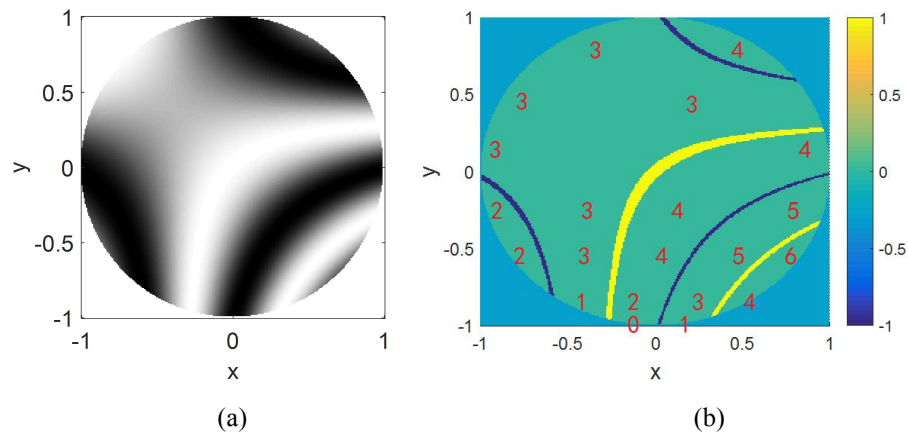

(a) (b)

**Fig.5** Parity of $m_{i,j}$ in different regions in the interferogram: (a)single shot fringe pattern and (b)$n_{i,j}$ distribution

Correspondingly, figure 6 shows the simulation results, and wavefront is in unit of rad. Figure 6(a) shows the original wavefront. Its PV and rms of the original wavefront are 1.01λ and 0.211λ, respectively. The estimated phase $\Delta\phi_\pi$ from figure 5(a) is shown in figure 6(b). The final estimated one is shown in figure 6(c). And its PV and rms are 1.01λ and 0.211λ, respectively. And the residual wavefront is shown in figure 6(d).  And its PV and rms are 0.00054λ and 0.000125λ, respectively. Therefore, the precision of the wavefront recovery is theoretically very high, and the simulated error rms is less than one thousandth of a wavelength.

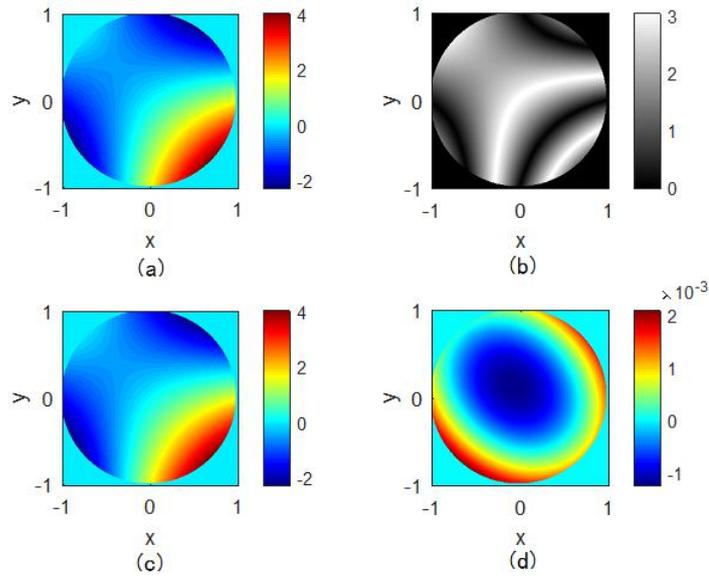

**Fig.6** Single shot interferogram and its extracted phase:(a)original wavefront in rad, (b)D$f_\pi$ in rad, (c) extracted wavefront in rad and (d) wavefront error in rad.

As a comparison, the results obtained with FFT method [28] are shown in the figure 7. An optimized spatial frequency is adopted and correspondent Fourier spectra were shown in figure 7(a). And only left side peak is used for phase extraction. It should be noted that the tilt and tip components in the original wavefront can't be extracted. Therefore, only high order aberrations without tilt and tip were estimated as shown in figure 7(b). Compared the extracted wavefront with the original one without tilt and tip terms, we could obtain the error wavefront. And PV and rms of residual wavefront are 0.5516$\lambda$ and 0.0169$\lambda$, respectively. Obviously, our method has a much higher precision.

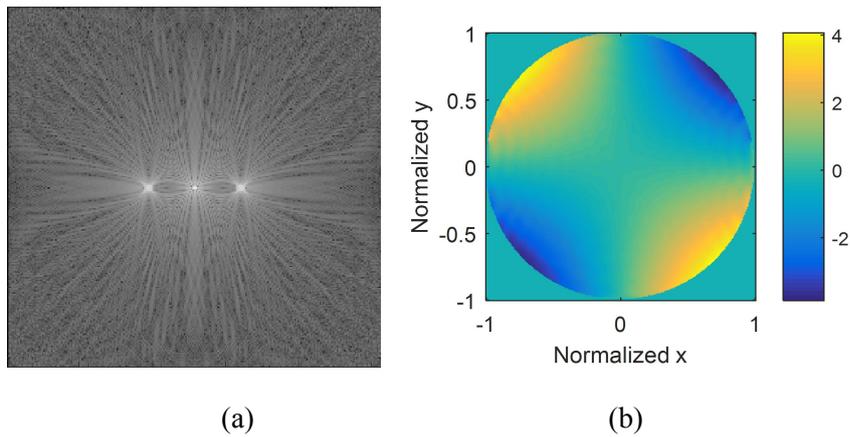

**Fig.7** Results of FFT method: (a)Fourier spectra and (b) extracted wavefront, unit: rad.

In fact, there are two main factors that will deteriorate the precision of phase extraction. First, it is the noise in fringe patterns. It should be noted that the fringe patterns used in theoretical simulation do not include noise. Therefore, it will be very easy to find the peak and vale traces in the fringe pattern. Noise will lead to some fake peak or vale points, which will make the peak and vale trace very complex. Therefore, filtering is necessary for an experimental fringe pattern. Second, it is easy to introduce error at the pixels on the peak or vale trace as shown in figure 5(b). Therefore, Zernike polynomial fitting is valuable.

## 4 Experimental interference fringes and phase recovery

The interferogram acquired in lab is shown in figure 8(a). Fringe normalization and filtering are important because of the noise in the fringe data as shown in figure 8(a). Therefore, necessary image processing is valuable so that the processed fringe pattern has a high signal to noise ratio(SNR). A high SNR image is key to improve the accuracy of phase extracting. Gray map gradient calculation is also used to find the peak and vale traces accurately in the interference fringe pattern, and then $n_{i,j}$ for every pixels is calculated. From the interference fringe in figure 8(a), the wavefront to be measured is calculated as shown in figure 8(b). The unit in the wave is radians, and correspondingly PV and rms are 4.47rad and 1.15rad, respectively.

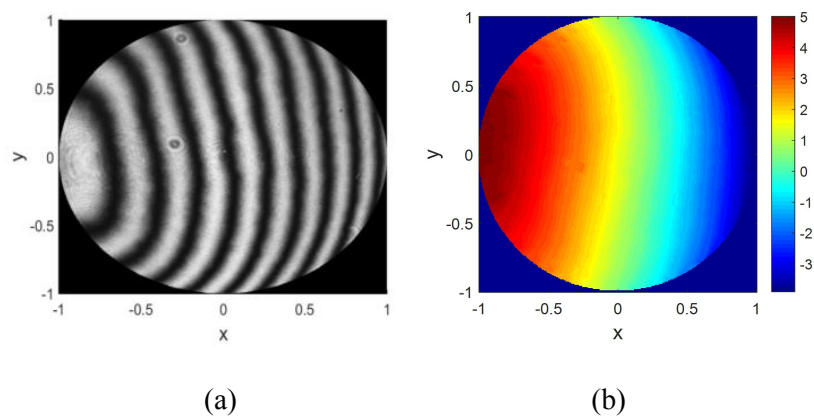

(a)          (b)

**Fig.8** Interferogram obtained by experimental measurement and recovered wavefront: (a)Subsectioa) interference fringe and (b) wavefront, unit: rad.

## 5 Conclusion

In the paper, we presented a simple and effective method theoretically and experimentally to evaluate the wavefront with a single shot fringe pattern. The theoretical simulation results show that the error is very small with PV of 0.00054λ, and rms of 0.000125λ, which are much better than those obtained with traditional FFT method. During large aperture aspheric surface porcessing and testing, we could use the optical layout as shown in figure 4 to get the prior knowledge. Then, it will be helpful for us to remove GSA and determine the reasonable and accurate wavefront of the optical element. In this case, the inclination direction of the final surface to be tested or the concave and convex direction of the curved surface can be judged reasonably. The proposed method has the following advantages: First, the hardware requirements are low: no expensive and high precision phase shifter in the traditional phase shifting interferometer is necessary; Secondly, the calculation is simple. Complex calculations such as Fourier transform and Hilbert transform are not necessary; Finally, it is effective for a wide range of situations in many applications. On the contrary, the Fourier transform method requires carrier frequency. The Hilbert transform method requires special processing for a close the interference fringe pattern. Our method does not have these limitations and is also effective for complex interferogram with closed and/or open fringe patterns in applications. The presented phase evaluation method based on single shot fringe pattern could also be helpful for other applications.

*Acknowledgments*

This work was supported by National Natural Science Foundation of China (61475152) and Jiangsu Six talent peaks program(GDZB-011).